\documentclass[letter]{aa}

\usepackage{graphicx}
\usepackage{array}
\usepackage{txfonts}
\usepackage{caption}
\usepackage{lipsum}
\usepackage{hyperref}
\hypersetup{colorlinks, allcolors=blue}
\usepackage{url}
\usepackage{natbib}
\usepackage{breqn}
\usepackage{diagbox}
\usepackage{cuted} 
\usepackage{lscape}
\usepackage[mathscr]{euscript}
\usepackage{placeins}
\usepackage{titlesec}

\newcommand{\correction}[1]{{ \color{red} #1}}
\renewcommand{\correction}[1]{{#1}}

\begin{document} 
   \title{The late Miocene $^{10}$Be anomaly and the possibility of a supernova}
\author{  E. Maconi\inst{1},
          J. Alves\inst{1},
          J. Großschedl\inst{2},
          A. Rottensteiner\inst{1},
          C. Swiggum\inst{1},
          \and
          S. Ratzenb\"ock\inst{3}
          }
\institute{
        University of Vienna, Department of Astrophysics, T\"urkenschanzstraße 17, 1180 Wien, Austria\\ \email{efrem.maconi@univie.ac.at}
        \and
        Astronomical Institute of the Czech Academy of Sciences, Boční II 1401, 141 31 Prague 4, Czech Republic
        \and
        Center for Astrophysics | Harvard \& Smithsonian, 60 Garden St., Cambridge, MA 02138, USA
        }

\date{Received ...; accepted ...}

\titlerunning{The $^{10}$Be anomaly and the possibility of a supernova}
\authorrunning{E.~Maconi et al.}

\abstract 
{Recent measurements of cosmogenic $^{10}$Be in deep-ocean ferromanganese crusts from the central and northern Pacific have revealed an anomalous concentration between 11.5 and 9.0 Myr ago, peaking at 10.1 Myr. One possible explanation is a nearby supernova (SN) event. Motivated by this and by the proximity of the Solar System to the Orion star-forming region during that period, we estimated the probability that at least one SN occurred between the onset and peak of the anomaly. Using an open cluster catalog based on \mbox{\textit{Gaia} DR3}, we traced back the orbits of 2725 clusters and the Sun over the past 20 Myr and computed the expected number of SN events. We found 19 clusters with a probability greater than 1\% each of producing at least one SN within 100\,pc of the Sun in the time interval 11.5–10.1 Myr ago. The total cumulative probability exceeds zero at 35\,pc from the Sun and increases rapidly with distance, reaching 68\% near 100\,pc. Two young clusters dominate the SN probability: ASCC\,20 contributes most within 70\,pc, while OCSN\,61 becomes more significant beyond that distance.
Our results support the \correction{possibility} of an SN origin for the $^{10}$Be anomaly and highlight the importance of additional $^{10}$Be records from independent terrestrial archives to determine whether the anomaly is of astrophysical or terrestrial origin.}


   \maketitle
%

\section{Introduction}

The Solar System orbits the center of the Milky Way, together with billions of other stars and vast reservoirs of interstellar gas. Characterizing past environments crossed by the Sun, including possible encounters with large-scale Galactic structures or nearby supernovae (SNe), helps to find possible connections between Galactic environments and Earth's geological records, fostering interdisciplinary research \citep[e.g.,][]{Fuchs2006,Breitschwerdt2016,Koll2019,Wallner2021,Miller2022,Opher2024,Maconi2025,Zucker2025}.

From an astronomical perspective, the European Space Agency's \textit{Gaia} mission \citep{Gaia2016} has enabled significant advancements, revolutionizing our understanding of the local Galactic environment. For example, \textit{Gaia} data have been used to unveil the 3D structure of the solar neighborhood \citep[see e.g.,][]{Leike2020,Vergely2022,Edenhofer2024}, compile new molecular cloud catalogs \citep[see e.g.,][]{Zucker2019,Cahlon2024}, and significantly expand the open cluster census \citep[see e.g.,][]{Castro-Ginard2018,Cantat-Gaudin2018,Hunt2023}. 
These achievements have provided new insights into the formation and evolution of stellar clusters \citep[see e.g.,][]{Meingast2019,Swiggum2024}, helped constrain the structure and history of the Local Bubble \citep[][]{Zucker2022,ONeill2024}, and led to the identification of previously unknown Galactic structures, such as the Radcliffe wave \citep[][]{Alves2020}, along with their relation to the past trajectory of the Solar System \citep[see e.g.,][]{Maconi2025}.

Concurrent to these astronomical advancements, studies of long-lived radionuclides such as $^{60}$Fe ($t_{1/2}\sim2.60 \, \mathrm{Myr}$; \citealt{Rugel2009,Wallner2015-60FeLife}) in geological archives have revealed signatures indicative of nearby SNe or encounters with interstellar regions enriched with these elements \citep[see e.g.,][]{Knie1999,Wallner2016,Wallner2021,Koll2019}. 
In addition, cosmogenic nuclides such as $^{14}$C \correction{($t_{1/2} \sim 5.700 \, \mathrm{ky}$; \citealt{Kutschera2013})} and $^{10}$Be ($t_{1/2} \sim 1.39\,\mathrm{Myr}$; \citealt{Chmeleff2010,Korschinek2010}), are used for archaeological and geological dating on kiloyear to million-year timescales.
Anomalies in their concentration profiles are important both for their physical interpretation and as potential chronological anchor markers if found in multiple independent archives \citep[see e.g.,][]{Dee2016}.

Recently, \citet{Koll2025} reported on the discovery of a $^{10}$Be anomaly in deep ocean crusts of the central and northern Pacific during the late Miocene. The origin of this anomaly, dated between 11.5 and 9.0 Myr ago and peaking at 10.1 Myr, remains uncertain, and several scenarios have been discussed by the authors. As $^{10}$Be is produced by cosmic-ray (CR) spallation in the upper atmosphere \citep[see e.g.,][]{Webber2003}, one possibility is that a nearby SN may be responsible for the $^{10}$Be excess.

In a recent study, \citet{Maconi2025} showed that around 11.5 Myr ago, at the onset of the $^{10}$Be anomaly, the Solar System was exiting the Radcliffe wave, leaving behind the Orion star-forming region, \correction{where 10-20 SNe likely occurred over the past 12 Myr \citep[see e.g.,][]{Bally2008}.} Given the Sun's proximity at that time to several massive young clusters, a nearby SN event is a \correction{possible} explanation for the observed $^{10}$Be anomaly. 
In this work, we test this hypothesis by integrating the orbits of the Sun and a large sample of clusters back in time over the past 20 Myr and by estimating the probability that a SN occurred within a given distance of the Solar System during the $^{10}$Be anomaly. 

\section{Data}\label{sect:data}

We primarily used the open cluster catalog by \citet{Hunt2023}, which \correction{is based} on the \textit{Gaia} DR3 astrometric data \citep{Gaia2023} and contains a total of 7166 star clusters  \citep[see][for details]{Hunt2023}. 
We complemented this catalog with four additional clusters not included in \citet{Hunt2023}; namely, CWNU\,1028, NGC\,1977, OC\,0340, and UBC\,207. These clusters are part of the Radcliffe wave and were identified and used in previous studies \citep{Konietzka2024,Maconi2025}, as is detailed in Appendix~\ref{app:cls_PosVel_recomputation}. 
Moreover, we updated the stellar memberships of six clusters in the Orion region (ASCC\,19, ASCC\,20, OCSN\,56, OCSN\,61, OCSN\,65, and Theia\,13) with additional members identified using the significance mode analysis ({\tt SigMA}) clustering algorithm \citep[][]{Ratzenboech2023a}, applied specifically to the Orion region (A.~Rottensteiner, in preparation).

By cross-matching individual cluster members with supplementary radial velocity (RV) surveys, we were able to improve the accuracy of the mean kinematic data of the clusters over the \textit{Gaia}-only values. 
For our purposes, we only considered clusters with reliable RV measurements \mbox{(${e}_{\mathrm{RV}} < 5 \, \mathrm{km}\,\mathrm{s}^{-1}$)} available for at least three member stars per cluster. After imposing these kinematic quality criteria, our sample comprises 2725 clusters. Further details on the data curation are provided in Appendix~\ref{app:cls_PosVel_recomputation}.

\section{Methods}\label{sect:methods}

\subsection{Selection of the clusters of interest}\label{sect:methods-clsSelection}

We performed a preliminary orbital integration (see Sect.~\ref{sect:methods-orbitInt}) of our initial cluster sample, identifying those that approached the Solar System within a threshold distance of 200\,pc over the past 20\,Myr. This threshold was chosen to ensure completeness, considering that SNe occurring within 100-150\,pc may leave detectable traces on Earth \citep[see e.g.,][]{Fry2015}. This step reduced the sample to 278 clusters ($\sim$10\% of the initial sample).

For each of these 278 clusters, we performed 1000 orbital integrations, varying their initial positions and velocities by Monte Carlo (MC) sampling the uncertainty distributions. We then selected only the clusters that approached the Solar System within 100\,pc during the time span of the $^{10}$Be anomaly \citep[9–11.5 Myr ago; see][]{Koll2025}. 
We adopt 100\,pc as a conservative threshold, balancing the range at which SNe may leave traces on Earth with the need for proximity to explain the $^{10}$Be anomaly \citep{Koll2025}.
We do not explicitly account for the travel time between the SN event and the arrival of CRs at Earth, as this interval is negligible compared to the million-year-scale intervals studied here. This resulted in a subset of 71 clusters. 

\subsection{Age, mass, and SN estimation}\label{sect:methods-AgeMassSN}

For the subset of 71 selected clusters identified in Sect.~\ref{sect:methods-clsSelection}, we estimated ages, present-day masses, completeness-corrected masses, and the probability of SN events. 
For the age estimation, we used the {\tt Chronos}\footnote{{\tt Chronos} code: \href{https://github.com/sebastianratzenboeck/Chronos}{https://github.com/sebastianratzenboeck/Chronos}} {\tt Python} package \citep[see][]{Ratzenboech2023b}, which performs a Bayesian fit of theoretical isochrones to the color magnitude diagram of each cluster. We used the PARSEC isochrones models \citep{Bressan2012,Nguyen2022} and assumed solar metallicity ($Z_{\odot}=0.0158$) for the stars in the clusters.
We limited our fit to $M_G<10$ as stars fainter than this threshold have been empirically found to be less well described by isochrone models \citep[see e.g.,][]{Ratzenboech2023b,Rottensteiner2024}.

\begin{figure*}
    \centering
    \includegraphics[width=0.86\textwidth]{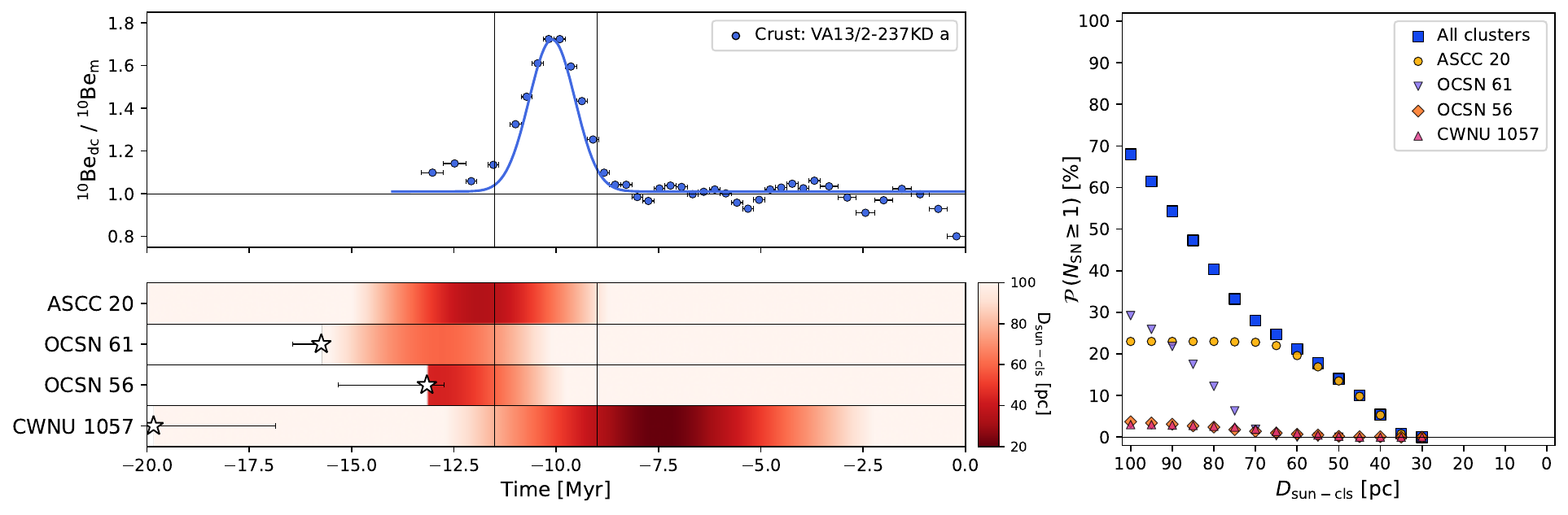}
    \caption{Overview of the $^{10}$Be anomaly, clusters proximity to the Solar System, and the associated SN probability. 
    \textit{Top left}: $^{10}$Be decay corrected profile ($^{10}$Be$_{\mathrm{dc}}$), normalized by the $^{10}$Be mean equilibrium surface concentration ($^{10}$Be$_{\mathrm{m}}$), for the crust VA13/2-237KD-a, as reported in Fig.~4 of \citet{Koll2025} (kindly provided in a processed form by the authors upon request). A Gaussian centered at 10.1 Myr, with a full width at half maximum of 1.4 Myr, is overplotted in blue. The two vertical black lines mark the onset and end of the anomaly.
    \textit{Bottom left}: Distance between the Sun and the four clusters discussed in Sect.~\ref{sect:results_discussion} over the past 20\,Myr. The color bar saturates at 100\,pc, the threshold distance adopted in this work for the SN probability study. If a cluster formed within this time interval, its formation time is marked by a star, and an error bar indicates the associated uncertainty.
    \textit{Right}: Probability of having at least one SN event between the onset and the peak of the anomaly (11.5–10.1 Myr ago) as a function of distance. The total probability for all clusters considered in this study, as well as for each of the four main clusters, is shown. Statistical errors for the data points are within 3\%. The numerical values corresponding to this figure are reported in Table~\ref{table_app:prob_atLeast1SN_allCLS}.
    }
    \label{fig:10BeAnomaly_DistSunCls_ProbAtLeast1SN}
\end{figure*}

The present-day cluster masses were estimated by adding up the masses of its stellar members, derived from the corresponding best-fit isochrone. 
We also computed the initial mass of the clusters corrected for incompleteness, as some star members may be missing due to observational limits, limitation of the clustering algorithm, and stellar evolution \citep[see e.g.,][]{Meingast2021,Hunt2023,Ratzenboech2023a}. Following \citet{Meingast2021}, we minimized the difference between the present-day mass function and a \citet{Kroupa2001} initial mass function (IMF). 
To account for the completeness limits of \textit{Gaia} data, we performed the minimization within the mass range $0.3-2 \, \mathrm{M}_{\odot}$. 
The low-mass end limit of the Kroupa IMF was set to $0.03 \, \mathrm{M}_{\odot}$ to account for objects below the hydrogen-burning limit, while no high-mass bound was imposed to account for massive stars potentially absent due to \textit{Gaia} brightness limit or past SNe.
The IMF was sampled using the {\tt IMF} {\tt Python} code\footnote{{\tt IMF} code: \href{https://github.com/keflavich/imf}{https://github.com/keflavich/imf}}. 

For each cluster, we then estimated the number of SNe that may have occurred while it was within 100\,pc of the Solar System and during the time interval between the onset and the peak of the $^{10}$Be anomaly (11.5-10.1\,Myr ago). This corresponds to the period during which a nearby SN would be most relevant to explain the peak in $^{10}$Be. 
For each of the 1000 orbital realizations (see Sect.~\ref{sect:methods-orbitInt}), we assigned an age and a mass to the cluster by sampling their respective posterior distributions as provided by {\tt Chronos}.
The sampled mass was used to generate a synthetic stellar population assuming a Kroupa IMF. 
The sampled age was used to compute the age of the cluster at the beginning and end of its proximity to the Sun within the time window of interest. 
We then counted the number of massive stars ($M>8\,\mathrm{M}_\odot$) with masses greater than the most massive star predicted by the PARSEC stellar evolutionary model for the corresponding ages. 
From this, we derived the probability of hosting at least one SN event during the considered time range and across different distance thresholds, both for each cluster individually and for the entire cluster ensemble.
This procedure was repeated 100 times to estimate the statistical uncertainties on the probabilities.

\subsection{Integration of the orbits}\label{sect:methods-orbitInt}

We computed the past orbital trajectories of the clusters and the Sun using the Galactic dynamics package {\tt galpy} \citep{Bovy2015}. This package enables the numerical orbit integration for different initial conditions (i.e., Galactocentric distance, solar height above the disk, velocity of the Sun, and velocity of the local standard of rest (LSR)) and various models for the Milky Way potential.

For this study, we adopted the {\tt MWPotential2014} model offered by {\tt galpy} as the gravitational potential of the Milky Way. This model includes a bulge, disk, and dark-matter halo component \citep[see][for details]{Bovy2015}.
We assumed a solar Galactocentric distance of $R_{\odot} = 8.33 \, \mathrm{kpc}$ \citep{Gillessen2009} and a solar height of $z_{\odot}= 27 \, \mathrm{pc}$ \citep{Chen2001}. The velocity of the LSR was set to $v_{\mathrm{LSR}} = 220 \, \mathrm{km}\,\mathrm{s}^{-1}$ and the velocity of the Sun relative to it to ($U_\odot,\, V_\odot,\, W_\odot$) = (11.1, 12.24, 7.25)$\,\mathrm{km}\,\mathrm{s}^{-1}$ \citep{Schoenrich2010}.
Orbital integrations were performed over the past 20 Myr with a 0.01 Myr time step using the {\tt dop853-c} Dormand–Prince integrator in {\tt galpy} for computational efficiency.

Statistical uncertainties in the positions and velocities of the clusters and of the Sun were addressed by repeating the orbital integration procedure 1000 times, each time using a new realization of the input data by MC sampling the associated uncertainties. 
In Appendix~\ref{app:initCondTest}, we assess the impact of adopting a different set of solar parameters and find that the relative distances of the Sun and clusters remain largely unchanged over the past 20 Myr, supporting the robustness of our conclusions.

\section{Results and discussion}\label{sect:results_discussion}

Various scenarios are considered by \citet{Koll2025} to explain the $^{10}$Be anomaly in the late Miocene. Some involve geological processes that could increase $^{10}$Be concentrations in ocean water without altering its atmospheric production rate, while others explore the possibility of an actual enhancement in Galactic CR flux reaching Earth. Among the most promising explanations are the onset and intensification of the Antarctic circumpolar current as a terrestrial cause, and either a nearby SN or the compression of the heliosphere by the passage of the Sun through a dense interstellar cloud as astrophysical origins. We refer the reader to their work for a more detailed discussion of all proposed scenarios and focus here solely on the SN hypothesis, further motivated by the fact that around the onset of the $^{10}$Be anomaly (11.5 Myr ago), the Solar System was leaving the Orion region of the Radcliffe wave behind \citep{Maconi2025}, and was therefore in proximity to an active star-forming region where several massive clusters were forming or had just formed. \correction{The difference from \citet{Maconi2025} is that here we consider all high-quality clusters in the solar neighborhood, rather than focusing only on those associated with the Radcliffe wave.}

We find that, out of the 2725 open clusters in our initial sample (see Sect.~\ref{sect:data}), 19 have a probability greater than 1\% of hosting at least one SN within 100\,pc of the Solar System and between the onset and the peak of the $^{10}$Be anomaly (11.5–10.1 Myr ago), and only four exceed this threshold within 70\,pc; namely, ASCC\,20, OCSN\,61, OCSN\,56, and CWNU\,1057. We estimate the total probability of at least one SN event within 35\,pc of the Sun to be around 1\%, rising to 5.4\% at 40\,pc, 14\% at 50\,pc, and 28\% at 70\,pc. At 100\,pc the total probability across all clusters reaches 68\%.
We note that none of the considered clusters come closer than 20\,pc, and thus none of them reach the critical distance (8–20\,pc) within which a SN could cause an extinction event \citep[see e.g.,][]{Gehrels2003,Thomas2023}.
In Fig.~\ref{fig:10BeAnomaly_DistSunCls_ProbAtLeast1SN}, we show the distances between the four mentioned clusters and the Solar System over the past 20 Myr, their individual SN probabilities, the total SN probability from all clusters during the relevant interval, and the $^{10}$Be concentration profile by \citet{Koll2025}.
In Table~\ref{table_app:prob_atLeast1SN_allCLS}, we list the 19 clusters of interest together with their SN probability for different threshold distances. Their properties, including ages and masses, are reported in Table~\ref{table_app:info_allCLS}, while their heliocentric positions and velocities are given in Table~\ref{table_app:pos_vel_allCLS}.

Among the 19 clusters, ASCC\,20 and OCSN\,61 (also known as OBP-b), both located in Orion, are the dominant contributors. ASCC\,20 is the only cluster with a significant contribution within 70\,pc (up to $\sim23$\%), while OCSN\,61 becomes increasingly relevant beyond 70\,pc, reaching 29\% at 100\,pc. 
For ASCC\,20, which reaches a minimum distance of $\sim34\,\mathrm{pc}$ from the Sun around 11.8\,Myr ago and remains within 100\,pc throughout the $^{10}$Be anomaly, we estimated an age of $\sim21.7\,\mathrm{Myr}$, consistent with values reported in the literature \citep[see e.g.,][]{Kos2019,Maconi2025}, and derive a present day mass of $\sim300\,\mathrm{M}_{\odot}$ and an incompleteness-corrected initial mass of $\sim500\,\mathrm{M}_{\odot}$. 
For OCSN\,61, which remains within 100\,pc between the onset and peak of the anomaly but never approaches the Sun closer than 60\,pc, we estimated an age of $\sim15.7\,\mathrm{Myr}$, a present-day mass of $\sim310\,\mathrm{M}_{\odot}$, and an incompleteness-corrected initial mass of $\sim540\,\mathrm{M}_{\odot}$.   
For both of these clusters, we updated the stellar membership using the {\tt SigMA} algorithm (see Sect.~\ref{sect:data} and Appendix~\ref{app:cls_PosVel_recomputation}), identifying new members compared to the one listed in \citet{Hunt2023}. 
Even with improved member selection, cluster catalogs are likely still incomplete, probably making our SN estimates conservative.
\correction{In Appendix~\ref{app:original_membership_lists}, we assess the impact of the membership lists on the SN probability estimation.}

To account for systematic uncertainties in the geological age dating, we adopt a conservative estimate of $\pm0.5$ Myr, based on the largest uncertainty reported for one of the samples (Crust-3) used by \citet{Koll2025} and previously analyzed by \citet{Wallner2021}. 
We repeated our analysis with two shifted time windows: [12.0, 10.6] Myr ago and [11.0, 9.6] Myr ago. We find that the total SN probability increases slightly in the older time window and decreases in the younger one, mainly due to changes in the proximity of ASCC\,20 and OCSN\,61 (see Fig.~\ref{fig_app:probAtLeast1SN_systematicErrorsStudy} and Appendix~\ref{app:systematic_errors_test}). These variations do not affect the result that a nearby SN remains a \correction{possible} explanation for the $^{10}$Be anomaly.

We note that the $^{10}$Be anomaly appears as a broad peak rather than a sharp one, as might be expected from an SN. However, the time interval derived from the crust might be broader than the actual signal, due to diffusion and redistribution processes. 
Moreover, \citet{Koll2025} consider the SN hypothesis a viable scenario only under certain energetic and geometrical conditions. We consider this \correction{possible} given the many uncertainties and physical processes involved in a SN and CR production, including the total SN energy \citep[see e.g.,][]{Kasen2009}, the efficiency of energy conversion into CR \citep[see e.g.,][]{Blasi2011}, as well as CR acceleration \citep[see e.g.,][]{Caprioli2012} and transport \citep[see e.g.,][]{Amato2018}. Future advancements in the understanding of these processes will help to better constrain the conditions under which an SN could produce such an anomaly.

The presence of other radionuclides could offer further insights into the origin of the $^{10}$Be anomaly and potentially support the nearby SN scenario.
For example, $^{60}$Fe, produced in SNe, has been detected in geological archives and linked to past SN events \citep{Wallner2021}. 
\correction{For the $^{10}$Be anomaly analyzed here, no concomitant $^{60}$Fe peak has been detected yet.} However, $^{10}$Be enhancements have not been observed in association with previously reported $^{60}$Fe peaks, and this apparent lack of correlation remains an open question. \correction{This may be explained by the distances of the SNe from the Solar System; for example, greater than 80\,pc for one of the $^{60}$Fe anomalies \citep{Breitschwerdt2016, Schulreich2023}.} Additionally, the detection of $^{60}$Fe within the time window considered here is further complicated by its advanced decay and the expected low concentrations.
Among other cosmogenic radionuclides, $^{53}$Mn ($t_{1/2} \sim 3.7\,\mathrm{Myr}$) is a valuable candidate for complementary investigation alongside $^{10}$Be, and future accelerator mass spectrometry facilities may enable its detection in the relevant time window \citep[see e.g.,][]{Koll2025}.

In conclusion, we find that a nearby SN remains a \correction{possible} explanation for the $^{10}$Be anomaly, especially given the Solar System’s proximity to the Orion region during that period. The estimated SN probability is nonzero at 35\,pc and increases with distance, with ASCC\,20 and OCSN\,61 emerging as the most promising candidate clusters. ASCC\,20 is the primary contributor up to 70\,pc, while OCSN\,61 becomes more relevant beyond that distance. Future investigations of $^{10}$Be records from terrestrial archives outside the Pacific ocean will be crucial to determine whether the observed anomaly reflects a global signal or a regional effect confined to this basin, helping to constrain its terrestrial or astrophysical origin.

\begin{acknowledgements}
\correction{We thank the referee A.\,Wallner for the insightful comments, which have improved the quality and readability of the paper.}
EM, JA, and AR were co-funded by the European Union (ERC, ISM-FLOW, 101055318). JG was co-funded by the European Union, Central Bohemian Region, and Czech Academy of Sciences, as part of the MERIT fellowship (MSCA-COFUND Horizon Europe, Grant agreement 101081195). This work has made use of data from the European Space Agency (ESA) mission \textit{Gaia} \href{https://www.cosmos.esa.int/gaia}{https://www.cosmos.esa.int/gaia}, processed by the \textit{Gaia} Data Processing and Analysis Consortium (DPAC,
\href{https://www.cosmos.esa.int/web/gaia/dpac/consortium}{https://www.cosmos.esa.int/web/gaia/dpac/consortium}). Funding for the DPAC has been provided by national institutions, in particular the institutions participating in the \textit{Gaia} Multilateral Agreement.
\end{acknowledgements}

%
%
\bibliographystyle{aa} 
\bibliography{references} 

\begin{appendix} 
\onecolumn

\section{Cluster catalog data}\label{app:cls_PosVel_recomputation}

In this section, we describe the compilation of open clusters used in this study and explain how we refined the positional and kinematic data obtained from the primary catalog.
As outlined in Sect.~\ref{sect:data}, we primarily used the open cluster catalog by \citet{Hunt2023}, which currently represents the largest open cluster search conducted homogeneously. This catalog was constructed using the Hierarchical Density-Based Spatial Clustering of Applications with Noise routine \citep[HDBSCAN;][]{McInnes2017} on \textit{Gaia} DR3 astrometric data \citep{Gaia2023}, combined with a statistical density test and a Bayesian convolutional neural network for result validation \citep[see][for details]{Hunt2023}. The catalog contains a total of 7166 star clusters.
Following \citet{Maconi2025}, we added four additional clusters that are associated with the Radcliffe wave \citep[see also][]{Konietzka2024}: CWNU\,1028, NGC\,1977, OC\,0340, and UBC\,207.

For six clusters in the Orion region (ASCC\,19, ASCC\,20, OCSN\,56, OCSN\,61, OCSN\,65, and Theia\,13), we were able to update the stellar membership using the {\tt SigMA} clustering algorithm \citep[Significance Mode Analysis;][]{Ratzenboech2023a}, applied to \textit{Gaia} data and specifically tuned for the Orion complex (A.~Rottensteiner, in preparation). 
The updated stellar membership for these six Orion clusters, compared to \citet{Hunt2023}, changes as follows: ASCC\,19 from 61 to 796, ASCC\,20 from 194 to 525, OCSN\,61 from 147 to 530, OCSN\,65 from 70 to 534, OCSN\,56 from  88 to 52, and Theia\,13 from 249 to 221. The latter two have slightly fewer stellar members compared to \citet{Hunt2023}. 
\correction{The fact that the SigMA-selected clusters are generally richer than those identified by \citet{Hunt2023}} can be explained by the different selection strategies. The latter study, which focuses on a larger stellar population sample, applies more stringent criteria and prioritizes precision over completeness to minimize the number of false positives associated with each cluster and to construct a uniform catalog across the entire Solar neighborhood. In contrast, the updated stellar memberships presented here are derived by applying {\tt SigMA} specifically to the Orion region, \correction{allowing for a more detailed analysis.}
We refer to \citet{Ratzenboech2023a} for a discussion on systematic biases coming from various clustering methods applied to the same region. 
The initial cluster sample comprises a total of 7170 members.
\correction{In Appendix~\ref{app:original_membership_lists}, we evaluate the impact of the membership lists on the SN probability estimation.}

To obtain more accurate 3D space motions for each cluster, we supplement the \textit{Gaia} DR3 RVs with additional RV data from supplementary surveys, including APOGEE-2 DR17 \citep{Abdurrouf2022}, GALAH DR4 \citep{Buder2025}, RAVE DR6 \citep{Steinmetz2020}, \textit{Gaia} ESO DR6 \citep{Randich2022}, two RV compilations \citep{Gontcharov2006,Torres2006}, \correction{and LAMOST DR10 or} LAMOST DR5 \citep{Zhao2012, Cui2012LAMOST} (DR5 as corrected by \citealt{Tsantaki2022} from the so called SoS catalog. SoS-LAMOST-DR5 was only used if the source was not in LAMOST DR10). 
In cases where a star has multiple RV measurements from different surveys, we selected the RV value with the lowest uncertainty. Additionally, we only include sources with RV errors smaller than $5 \, \mathrm{km}\,\mathrm{s}^{-1}$. To remove outliers, we applied 3-sigma clipping around each cluster's median RV value. We then derived the \correction{Heliocentric Galactic} Cartesian velocities ($U,\,V,\,W$) \correction{[km/s]} using the sub-sample of stars with valid RV measurements. For our study, we used medians in $XYZ$ and $UVW$, along with the associated uncertainties, as an estimate for the cluster's bulk position and motion. Only clusters with at least three stars with RV measurements are included in our final sample, reducing the initial sample from 7170 to 2725.

\section{Cluster properties}\label{app:cls_properties}

In this Section, we report the Tables that summarize the properties of the clusters as derived in the main part of the paper. The tables regard the 19 clusters that have a probability greater than 1\% of hosting an SN within 100\,pc of the Solar System and during the time interval 11.5–10.1\,Myr ago, as described in Sects.~\ref{sect:methods} and \ref{sect:results_discussion}.

Table~\ref{table_app:pos_vel_allCLS} lists the Heliocentric Galactic Cartesian positions and velocities.
Table~\ref{table_app:info_allCLS} provides various information for the clusters, including ages, masses, minimum Sun–cluster distances, and time intervals during which they remain within 100\,pc.
Table~\ref{table_app:prob_atLeast1SN_allCLS} reports the SN probabilities for different threshold distances between 100\,pc and 30\,pc.
\FloatBarrier
\newpage

\begin{table*}[h!]
\caption{Heliocentric Galactic Cartesian positions ($X,\,Y,\,Z$) and velocities ($U,\,V,\,W$), together with the corresponding standard errors of the mean, for the 19 clusters of interest.}  
\label{table_app:pos_vel_allCLS}
\centering                        
\resizebox{\textwidth}{!}{%
\begin{tabular}{lcccccccccccc}     
\hline\hline 
Name & $X$ & $Y$ & $Z$ & $U$ &$V$ & $W$ & $X_{\mathrm{err}}$ & $Y_{\mathrm{err}}$ & $Z_{\mathrm{err}}$ & $U_{\mathrm{err}}$ &$V_{\mathrm{err}}$ & $W_{\mathrm{err}}$ \\
 & $\mathrm{[pc]}$ & $\mathrm{[pc]}$ & $\mathrm{[pc]}$ & $\mathrm{[km\,s^{-1}]}$ &$\mathrm{[km\,s^{-1}]}$ & $\mathrm{[km\,s^{-1}]}$ & $\mathrm{[pc]}$ & $\mathrm{[pc]}$ & $\mathrm{[pc]}$ & $\mathrm{[km\,s^{-1}]}$ &$\mathrm{[km\,s^{-1}]}$ & $\mathrm{[km\,s^{-1}]}$ \\
\hline 
ASCC\,20 & -319.65 & -129.48 & -108.33 & -27.03 & -8.87 & -9.34 & 0.98 & 0.42 & 0.43 & 0.27 & 0.12 & 0.10 \\
ASCC\,24 & -164.72 & -124.71 & -26.95 & -13.65 & -9.72 & -10.00 & 0.93 & 0.97 & 0.83 & 0.61 & 0.39 & 0.19 \\
CWNU\,1057 & -114.40 & -63.81 & -54.58 & -13.45 & -9.39 & -7.95 & 0.62 & 1.38 & 0.56 & 1.10 & 0.61 & 0.57 \\
CWNU\,1111 & -256.25 & -56.66 & -42.29 & -19.22 & -9.50 & -9.33 & 1.74 & 1.03 & 0.41 & 1.67 & 0.44 & 0.25 \\
HSC\,1340 & -112.20 & 5.24 & -39.43 & -13.89 & -6.32 & -9.90 & 0.56 & 0.83 & 0.50 & 0.53 & 0.08 & 0.20 \\
HSC\,1373 & -107.87 & 29.02 & -82.49 & -12.43 & -5.41 & -5.95 & 1.46 & 2.50 & 0.83 & 0.91 & 0.29 & 0.99 \\
HSC\,1469 & -122.45 & -17.54 & -128.58 & -4.72 & -6.11 & -7.85 & 1.58 & 1.25 & 0.52 & 0.84 & 0.16 & 0.94 \\
HSC\,1523 & -257.27 & -56.94 & -75.51 & -16.28 & -9.18 & -6.13 & 1.68 & 1.03 & 0.66 & 2.04 & 0.43 & 0.60 \\
HSC\,1640 & -170.42 & -85.42 & -135.24 & -6.83 & -8.63 & -6.01 & 2.29 & 1.39 & 1.15 & 0.88 & 0.36 & 0.56 \\
HSC\,1687 & -312.44 & -208.38 & -46.00 & -23.38 & -24.10 & -6.82 & 1.48 & 2.16 & 0.74 & 0.18 & 0.16 & 0.18 \\
HSC\,1692 & -197.27 & -137.93 & -101.25 & -22.99 & -5.87 & -6.85 & 3.75 & 2.58 & 1.65 & 0.65 & 0.30 & 0.48 \\
NGC\,2232 & -258.61 & -177.32 & -41.08 & -20.31 & -12.89 & -10.70 & 0.83 & 0.61 & 0.21 & 0.49 & 0.33 & 0.08 \\
OCSN\,50 & -175.67 & 22.91 & -71.36 & -14.80 & -5.66 & -5.85 & 2.04 & 1.63 & 0.53 & 0.95 & 0.18 & 0.44 \\
OCSN\,56 & -345.38 & -124.03 & -97.58 & -26.34 & -8.58 & -8.71 & 3.05 & 1.12 & 0.95 & 1.55 & 0.61 & 0.38 \\
OCSN\,61 & -344.72 & -159.75 & -115.31 & -25.59 & -11.69 & -10.95 & 1.08 & 0.56 & 0.41 & 0.41 & 0.18 & 0.13 \\
OCSN\,64 & -268.51 & -132.13 & -122.54 & -26.17 & -5.31 & -5.58 & 11.81 & 5.84 & 5.39 & 1.31 & 0.78 & 0.68 \\
Theia\,69 & -115.07 & -120.16 & -21.54 & -12.56 & -18.83 & -4.54 & 0.52 & 0.77 & 0.46 & 0.81 & 0.84 & 0.22 \\
Theia\,71 & -177.34 & 7.91 & -57.94 & -14.92 & -6.11 & -6.07 & 0.79 & 0.85 & 0.48 & 1.77 & 0.01 & 0.44 \\
Theia\,72 & -212.29 & -187.58 & -26.09 & -18.24 & -22.33 & -6.41 & 1.83 & 1.40 & 0.43 & 1.13 & 0.90 & 0.20 \\
\hline                                
\end{tabular}
}
\end{table*}

\begin{table*}[h!]
\caption{Properties for the 19 clusters of interest.}
\label{table_app:info_allCLS}
\renewcommand{\arraystretch}{1.5}
\centering
\resizebox{\textwidth}{!}{%
\begin{tabular}{llcccccccccc}
\hline\hline
Name & Other name & $N_{\mathrm{*}}$ & $N_{\mathrm{*,RV}}$ & $N_{\mathrm{*,\ge8\,M_\odot}}$ & Age & $M^{\mathrm{cls}}$ & $M^{\mathrm{cls}}_{\mathrm{corr}}$ & $D^{\mathrm{Sun-cls}}_{\mathrm{min}}$ & $t^{\mathrm{Sun-cls}}_{D_{\mathrm{min}}}$ & $t^{\mathrm{Sun-cls}}_{\le100\,\mathrm{pc}}$ & $\mathrm{P}^{\mathrm{Sun-cls}}_{\le100\,\mathrm{pc}}$\\
& & & & & $\mathrm{[Myr]}$ & $\mathrm{[M_\odot]}$ & $\mathrm{[M_\odot]}$ & [pc] & $\mathrm{[Myr]}$ & $\mathrm{[Myr]}$ & \%  \\
\hline
ASCC\,20 & - & 525 & 110 & $5_{-2}^{+2}$ & $21.7_{-0.4}^{+0.5}$ & 306 & $513_{-61}^{+103}$ & ${34.2}{\,\scriptstyle\pm\,2.0}$ & ${-11.9}{\,\scriptstyle\pm\,0.1}$ & [${-15.0}{\,\scriptstyle\pm\,0.1}, {-8.7}{\,\scriptstyle\pm\,0.1}$] & $100$ \\
ASCC\,24 & - & 18 & 4 & $0_{-0}^{+1}$ & $23.4_{-2.8}^{+4.7}$ & 24 & $24_{-9}^{+25}$ & ${82.9}{\,\scriptstyle\pm\,3.2}$ & ${-9.9}{\,\scriptstyle\pm\,0.2}$ & [${-13.1}{\,\scriptstyle\pm\,0.3}, {-6.9}{\,\scriptstyle\pm\,0.4}$] & $100$ \\
CWNU\,1057 & - & 54 & 14 & $0_{-0}^{+1}$ & $19.8_{-3.0}^{+3.3}$ & 25 & $46_{-15}^{+34}$ & ${20.5}{\,\scriptstyle\pm\,6.4}$ & ${-7.5}{\,\scriptstyle\pm\,0.3}$ & [${-12.7}{\,\scriptstyle\pm\,0.6}, {-2.3}{\,\scriptstyle\pm\,0.1}$] & $100$ \\
CWNU\,1111 & - & 40 & 8 & $0_{-0}^{+1}$ & $33.4_{-6.8}^{+2.9}$ & 25 & $59_{-16}^{+37}$ & ${95.7}{\,\scriptstyle\pm\,11.2}$ & ${-10.7}{\,\scriptstyle\pm\,0.5}$ & [${-12.3}{\,\scriptstyle\pm\,0.5}, {-8.7}{\,\scriptstyle\pm\,0.9}$] & $68$ \\
HSC\,1340 & - & 194 & 65 & $2_{-1}^{+1}$ & $24.0_{-1.8}^{+3.3}$ & 94 & $205_{-38}^{+61}$ & ${56.5}{\,\scriptstyle\pm\,2.1}$ & ${-5.6}{\,\scriptstyle\pm\,0.1}$ & [${-10.2}{\,\scriptstyle\pm\,0.1}, {-1.2}{\,\scriptstyle\pm\,0.1}$] & $100$ \\
HSC\,1373 & - & 46 & 10 & $0_{-0}^{+1}$ & $20.4_{-3.2}^{+2.8}$ & 19 & $39_{-11}^{+32}$ & ${76.4}{\,\scriptstyle\pm\,4.2}$ & ${-7.6}{\,\scriptstyle\pm\,0.4}$ & [${-11.7}{\,\scriptstyle\pm\,0.6}, {-3.5}{\,\scriptstyle\pm\,0.3}$] & $100$ \\
HSC\,1469 & - & 41 & 15 & $0_{-0}^{+1}$ & $23.0_{-1.3}^{+3.5}$ & 23 & $56_{-15}^{+34}$ & ${91.2}{\,\scriptstyle\pm\,8.5}$ & ${-11.7}{\,\scriptstyle\pm\,0.6}$ & [${-14.9}{\,\scriptstyle\pm\,1.3}, {-8.7}{\,\scriptstyle\pm\,1.1}$] & $85$ \\
HSC\,1523 & - & 17 & 4 & $0_{-0}^{+1}$ & $20.9_{-1.2}^{+3.3}$ & 13 & $40_{-13}^{+30}$ & ${89.2}{\,\scriptstyle\pm\,17.7}$ & ${-13.1}{\,\scriptstyle\pm\,0.9}$ & [${-15.5}{\,\scriptstyle\pm\,0.7}, {-10.1}{\,\scriptstyle\pm\,1.3}$] & $74$ \\
HSC\,1640 & Eridanus-North & 128 & 24 & $1_{-1}^{+1}$ & $15.7_{-1.9}^{+0.1}$ & 63 & $138_{-27}^{+53}$ & ${93.5}{\,\scriptstyle\pm\,12.5}$ & ${-14.7}{\,\scriptstyle\pm\,0.5}$ & [${-17.8}{\,\scriptstyle\pm\,1.2}, {-11.8}{\,\scriptstyle\pm\,1.1}$] & $69$ \\
HSC\,1687 & - & 19 & 6 & $0_{-0}^{+1}$ & $33.5_{-2.2}^{+3.8}$ & 14 & $35_{-12}^{+28}$ & ${95.8}{\,\scriptstyle\pm\,2.6}$ & ${-10.5}{\,\scriptstyle\pm\,0.1}$ & [${-11.4}{\,\scriptstyle\pm\,0.3}, {-9.7}{\,\scriptstyle\pm\,0.2}$] & $94$ \\
HSC\,1692 & - & 24 & 4 & $0_{-0}^{+1}$ & $29.9_{-2.0}^{+2.1}$ & 13 & $34_{-11}^{+26}$ & ${82.0}{\,\scriptstyle\pm\,4.6}$ & ${-9.9}{\,\scriptstyle\pm\,0.3}$ & [${-12.1}{\,\scriptstyle\pm\,0.5}, {-7.6}{\,\scriptstyle\pm\,0.3}$] & $100$ \\
NGC\,2232 & - & 286 & 101 & $3_{-2}^{+1}$ & $26.8_{-0.9}^{+1.0}$ & 171 & $293_{-47}^{+80}$ & ${92.2}{\,\scriptstyle\pm\,2.2}$ & ${-11.7}{\,\scriptstyle\pm\,0.2}$ & [${-13.2}{\,\scriptstyle\pm\,0.1}, {-10.1}{\,\scriptstyle\pm\,0.4}$] & $100$ \\
OCSN\,50 & - & 24 & 9 & $0_{-0}^{+1}$ & $16.3_{-2.2}^{+3.5}$ & 14 & $34_{-12}^{+30}$ & ${84.0}{\,\scriptstyle\pm\,4.2}$ & ${-10.0}{\,\scriptstyle\pm\,0.4}$ & [${-13.1}{\,\scriptstyle\pm\,0.4}, {-6.9}{\,\scriptstyle\pm\,0.6}$] & $100$ \\
OCSN\,56 & omega-Ori & 52 & 5 & $0_{-0}^{+1}$ & $13.2_{-0.4}^{+2.2}$ & 23 & $45_{-13}^{+32}$ & ${44.9}{\,\scriptstyle\pm\,8.6}$ & ${-13.0}{\,\scriptstyle\pm\,0.6}$ & [${-16.1}{\,\scriptstyle\pm\,0.7}, {-9.9}{\,\scriptstyle\pm\,0.6}$] & $100$ \\
OCSN\,61 & OBP-b & 530 & 102 & $5_{-2}^{+2}$ & $15.7_{-0.1}^{+0.7}$ & 312 & $540_{-71}^{+99}$ & ${59.2}{\,\scriptstyle\pm\,3.3}$ & ${-12.8}{\,\scriptstyle\pm\,0.1}$ & [${-15.5}{\,\scriptstyle\pm\,0.1}, {-10.1}{\,\scriptstyle\pm\,0.2}$] & $100$ \\
OCSN\,64 & OBP-e & 67 & 11 & $1_{-1}^{+0}$ & $22.6_{-1.8}^{+2.1}$ & 48 & $81_{-22}^{+43}$ & ${77.2}{\,\scriptstyle\pm\,12.0}$ & ${-11.4}{\,\scriptstyle\pm\,0.7}$ & [${-13.7}{\,\scriptstyle\pm\,1.1}, {-9.2}{\,\scriptstyle\pm\,0.6}$] & $97$ \\
Theia\,69 & - & 31 & 8 & $0_{-0}^{+1}$ & $24.7_{-0.9}^{+2.3}$ & 17 & $37_{-12}^{+30}$ & ${33.7}{\,\scriptstyle\pm\,5.4}$ & ${-6.9}{\,\scriptstyle\pm\,0.2}$ & [${-10.9}{\,\scriptstyle\pm\,0.4}, {-3.0}{\,\scriptstyle\pm\,0.1}$] & $100$ \\
Theia\,71 & - & 13 & 3 & $0_{-0}^{+1}$ & $14.4_{-3.6}^{+8.5}$ & 6 & $20_{-8}^{+25}$ & ${76.4}{\,\scriptstyle\pm\,8.3}$ & ${-9.7}{\,\scriptstyle\pm\,0.7}$ & [${-13.4}{\,\scriptstyle\pm\,0.5}, {-6.1}{\,\scriptstyle\pm\,1.0}$] & $99$ \\
Theia\,72 & - & 140 & 14 & $1_{-1}^{+1}$ & $30.7_{-1.7}^{+1.8}$ & 70 & $129_{-27}^{+47}$ & ${64.1}{\,\scriptstyle\pm\,9.2}$ & ${-9.2}{\,\scriptstyle\pm\,0.3}$ & [${-11.8}{\,\scriptstyle\pm\,0.4}, {-6.7}{\,\scriptstyle\pm\,0.4}$] & $100$ \\
\hline
\end{tabular}
}
\tablefoot{For each cluster, the following information is provided: \correction{name as given in its source catalog}, along with an alternative commonly used name if available; number of stellar members ($N_{\mathrm{*}}$); number of stars with RV ($N_{\mathrm{*,RV}}$); expected number of massive stars ($N_{\mathrm{*,\ge8\,M_\odot}}$); estimated isochronal age; cluster mass derived from the stellar members ($M^{\mathrm{cls}}$); initial cluster mass corrected for incompleteness ($M^{\mathrm{cls}}_{\mathrm{corr}}$); minimum Sun–cluster distance ($D^{\mathrm{Sun-cls}}_{\mathrm{min}}$); time of minimum approach ($t^{\mathrm{Sun-cls}}_{D_{\mathrm{min}}}$); time during which the Sun and the cluster are within 100\,pc ($t^{\mathrm{Sun-cls}}_{\le100\,\mathrm{pc}}$); and probability for the Sun and the cluster to be within 100 pc $\mathrm{P}^{\mathrm{Sun-cls}}_{\le100\,\mathrm{pc}}$. The expected number of massive stars was estimated using the initial cluster mass corrected for incompleteness and a synthetic stellar population assuming a Kroupa IMF.
}
\end{table*}

\begin{table*}[h!]
\caption{Probability of at least one SN event between 11.5 and 10.1 Myr ago as a function of distance, reported for all clusters combined and for each of the 19 clusters of interest.}
\label{table_app:prob_atLeast1SN_allCLS}    
\centering                        
\resizebox{\textwidth}{!}{%
\begin{tabular}{lccccccccccccccc}
\hline\hline              
Name & 100\,pc & 95\,pc & 90\,pc & 85\,pc & 80\,pc & 75\,pc & 70\,pc & 65\,pc & 60\,pc & 55\,pc & 50\,pc & 45\,pc & 40\,pc & 35\,pc & 30\,pc \\
\hline                      
All clusters & 68.0 & 61.5 & 54.3 & 47.3 & 40.3 & 33.2 & 28.0 & 24.7 & 21.2 & 17.8 & 13.9 & 10.0 & 5.4 & 0.8 & 0.0 \\
ASCC\,20 & \textbf{23.0} & \textbf{23.0} & \textbf{23.0} & \textbf{23.0} & \textbf{23.0} & \textbf{22.9} & \textbf{22.8} & \textbf{22.0} & \textbf{19.6} & \textbf{16.9} & \textbf{13.5} & \textbf{9.8} & \textbf{5.3} & \textbf{0.8} & 0.0 \\
ASCC\,24 & 1.9 & 1.9 & 1.8 & 1.0 & 0.1 & - & - & - & - & - & - & - & - & - & - \\
CWNU\,1057 & 3.0 & 3.0 & 2.9 & \textbf{2.8} & \textbf{2.6} & \textbf{2.4} & \textbf{1.9} & \textbf{1.4} & \textbf{0.9} & \textbf{0.5} & \textbf{0.2} & \textbf{0.1} & 0.0 & 0.0 & 0.0 \\
CWNU\,1111 & 1.5 & 1.2 & 0.7 & 0.3 & 0.1 & 0.0 & 0.0 & - & - & - & - & - & - & - & - \\
HSC\,1340 & 1.1 & 0.1 & 0.0 & 0.0 & 0.0 & 0.0 & 0.0 & 0.0 & 0.0 & 0.0 & - & - & - & - & - \\
HSC\,1373 & 2.4 & 1.9 & 1.1 & 0.4 & 0.1 & 0.0 & 0.0 & 0.0 & - & - & - & - & - & - & - \\
HSC\,1469 & 2.6 & 1.9 & 1.2 & 0.5 & 0.2 & 0.0 & 0.0 & - & - & - & - & - & - & - & - \\
HSC\,1523 & 1.2 & 1.0 & 0.8 & 0.6 & 0.4 & 0.3 & 0.2 & 0.1 & 0.1 & 0.0 & 0.0 & - & - & - & - \\
HSC\,1640 & 1.6 & 0.7 & 0.3 & 0.1 & 0.0 & 0.0 & 0.0 & 0.0 & 0.0 & - & - & - & - & - & - \\
HSC\,1687 & 1.2 & 0.4 & 0.0 & - & - & - & - & - & - & - & - & - & - & - & - \\
HSC\,1692 & 1.5 & 1.4 & 1.2 & 0.7 & 0.3 & 0.1 & 0.0 & - & - & - & - & - & - & - & - \\
NGC\,2232 & \textbf{9.7} & \textbf{5.5} & 0.7 & - & - & - & - & - & - & - & - & - & - & - & - \\
OCSN\,50 & 3.2 & 3.2 & 2.8 & 1.4 & 0.2 & 0.0 & - & - & - & - & - & - & - & - & - \\
OCSN\,56 & 3.7 & 3.4 & \textbf{3.1} & 2.7 & 2.4 & 1.8 & 1.4 & \textbf{1.0} & \textbf{0.7} & \textbf{0.5} & \textbf{0.2} & \textbf{0.1} & \textbf{0.1} & 0.0 & 0.0 \\
OCSN\,61 & \textbf{29.2} & \textbf{25.9} & \textbf{21.8} & \textbf{17.5} & \textbf{12.2} & \textbf{6.3} & \textbf{1.9} & 0.3 & 0.0 & 0.0 & 0.0 & - & - & - & - \\
OCSN\,64 & 3.9 & 3.6 & \textbf{3.1} & 2.5 & 1.7 & 1.0 & 0.5 & 0.3 & 0.1 & 0.0 & 0.0 & 0.0 & - & - & - \\
Theia\,69 & 1.3 & 1.0 & 0.7 & 0.4 & 0.2 & 0.1 & 0.0 & 0.0 & 0.0 & 0.0 & 0.0 & 0.0 & 0.0 & 0.0 & 0.0 \\
Theia\,71 & 2.1 & 2.1 & 1.9 & 1.8 & 1.1 & 0.4 & 0.0 & 0.0 & 0.0 & - & - & - & - & - & - \\
Theia\,72 & 4.4 & 4.1 & 3.7 & 3.0 & 2.3 & 1.5 & 0.8 & 0.3 & 0.1 & 0.0 & 0.0 & 0.0 & - & - & - \\
\hline                                 
\end{tabular}
}
\tablefoot{For each distance, the three highest probabilities are highlighted in bold. Dashes indicate that the Solar System did not reach that distance from the cluster.}
\end{table*}

\section{Initial condition test}\label{app:initCondTest}

We tested the robustness of our result by varying the initial solar parameters (i.e., the Sun's height above the disk, Galactocentric distance, and velocity with respect to the LSR), as there is no unique definition for these quantities. We then integrated the orbits of the Solar System and the clusters and performed the same analysis described in Sect.~\ref{sect:methods}. 

For this test, we assumed a Galactocentric distance of $R_{\odot} = 8.122 \, \mathrm{kpc}$ \citep{GRAVITY2018}, a vertical height of $z_{\odot}= 20.8 \, \mathrm{pc}$ \citep{Bennett2019}, and a solar velocity relative to the LSR of ($U_\odot,\, V_\odot,\, W_\odot$) = (11.1, 12.24, 7.25)$\,\mathrm{km}\,\mathrm{s}^{-1}$ \citep{Schoenrich2010}.
We find that the past relative distances of the Sun and the clusters, and thus the resulting probabilities of having an SN within the $^{10}$Be anomaly window, remain largely unchanged over the past 20 Myr. This can be explained by the relatively short integration time and by the fact that we consider the relative distances between the clusters and the Sun rather than their absolute positions. This is also in agreement with previous studies that used similar integration periods \citep[see e.g.,][]{Miret-Roig2020}, further supporting the robustness of our conclusions.

\FloatBarrier
\section{Systematic error study}\label{app:systematic_errors_test}

In this section, we assess the impact of the systematic uncertainties in the geological age dating on our results. 
A shift in the timing of the $^{10}$Be anomaly would alter the interval during which the Solar System is close to specific clusters, potentially affecting the SN probability estimates.
To test the impact of this on our results, we adopted a systematic uncertainty on the geological age dating of $\pm0.5\,\mathrm{Myr}$. 
This value is the largest systematic uncertainty we were able to find for the geological samples used in the paper by \citet{Koll2025} and is sourced from the work by \citet{Wallner2021} on Crust-3, one of the samples also used in \citet{Koll2025}. We then repeated the SN probability analysis (see Sect.~\ref{sect:methods-AgeMassSN}) using two shifted time windows for the $^{10}$Be anomaly: [12.0, 10.6] Myr and [11.0, 9.6] Myr.

\begin{figure*}
\sidecaption
  \includegraphics[width=12cm]{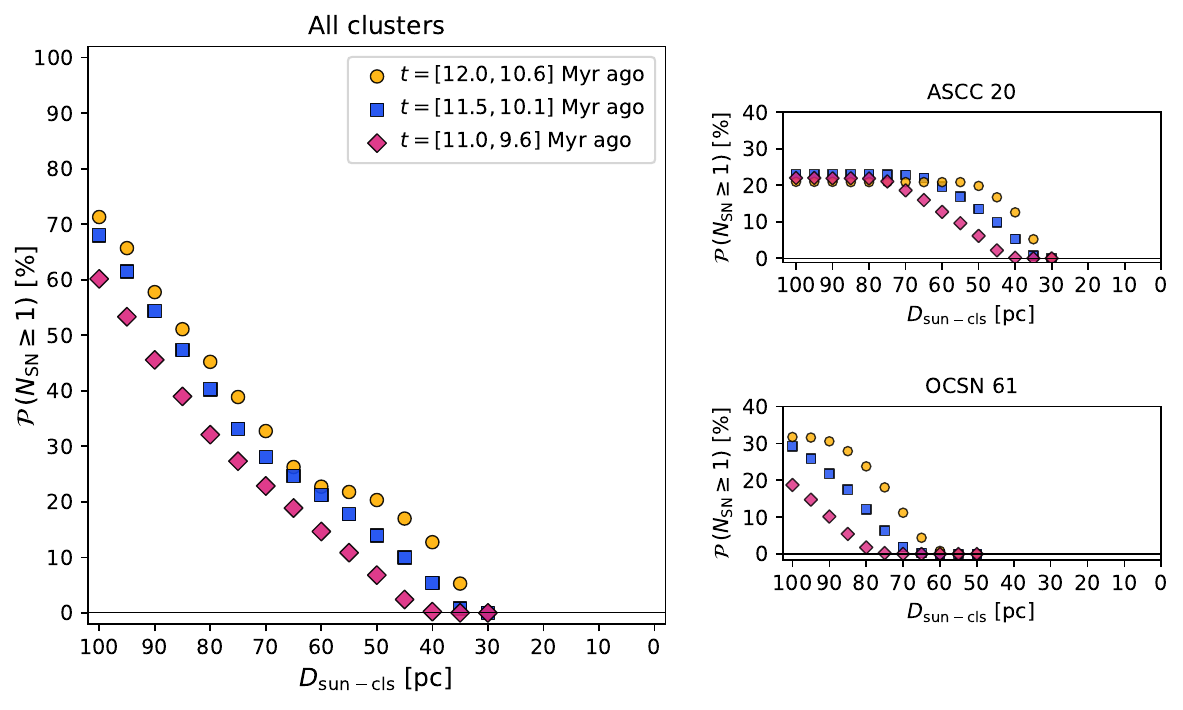}
     \caption{Probability of having at least one SN event as a function of distance, evaluated over three different $^{10}$Be anomaly time windows to assess the impact of systematic uncertainties on geological age dating. Blue squares correspond to the time window used in this study ([11.5, 10.1] Myr ago), based on \citet{Koll2025}. Yellow circles represent the older time window ([12.0, 10.6] Myr ago), and purple diamonds the younger one ([11.0, 9.6] Myr ago). The left panel shows the total probability derived from all clusters. The right panel shows the probabilities for ASCC\,20 (top) and OCSN\,61 (bottom).}
     \label{fig_app:probAtLeast1SN_systematicErrorsStudy}
\end{figure*}

We find that the total SN probability is only weakly affected by the assumed $\pm0.5\,\mathrm{Myr}$ systematic age uncertainty. 
As shown in the left panel of Fig.~\ref{fig_app:probAtLeast1SN_systematicErrorsStudy}, the total SN probability slightly increases for the older time window and slightly decreases for the younger one, primarily due to changes in the proximity of ASCC\,20 and OCSN\,61. 
The two main contributors are still ASCC\,20 and OCSN\,61. The right panel of Fig.~\ref{fig_app:probAtLeast1SN_systematicErrorsStudy} shows how the time shift affects the SN probability for these two clusters.
These results confirm that our main conclusion remains robust: a nearby SN is a viable explanation for the $^{10}$Be anomaly.

\correction{
\FloatBarrier
\section{Probability using the original cluster stellar memberships}\label{app:original_membership_lists}
}

In this Section, we present the results for the SN probability computed using the original stellar memberships for the Orion clusters.
As described in Sect.~\ref{sect:methods} and in the Appendix~\ref{app:cls_PosVel_recomputation}, our main analysis is based on the catalog by \citet{Hunt2023}, to which we:
\begin{itemize}
    \item added the four additional clusters CWNU\,1028, NGC\,1977, OC\,0340, and UBC\,207;
    \item updated the stellar membership list of the Orion clusters ASCC\,19 (from 61 to 796), ASCC\,20 (from 194 to 525), OCSN\,56 (88 to 52), OCSN\,61 (147 to 530), OCSN\,65 (70 to 534), and Theia\,13 (249 to 221). Among these clusters, some of them (e.g., ASCC\,20, OCSN\,56, OCSN\,61) have a significant contribution to the total SN probability. 
\end{itemize}

\begin{figure*}
\sidecaption
  \includegraphics[width=12cm]{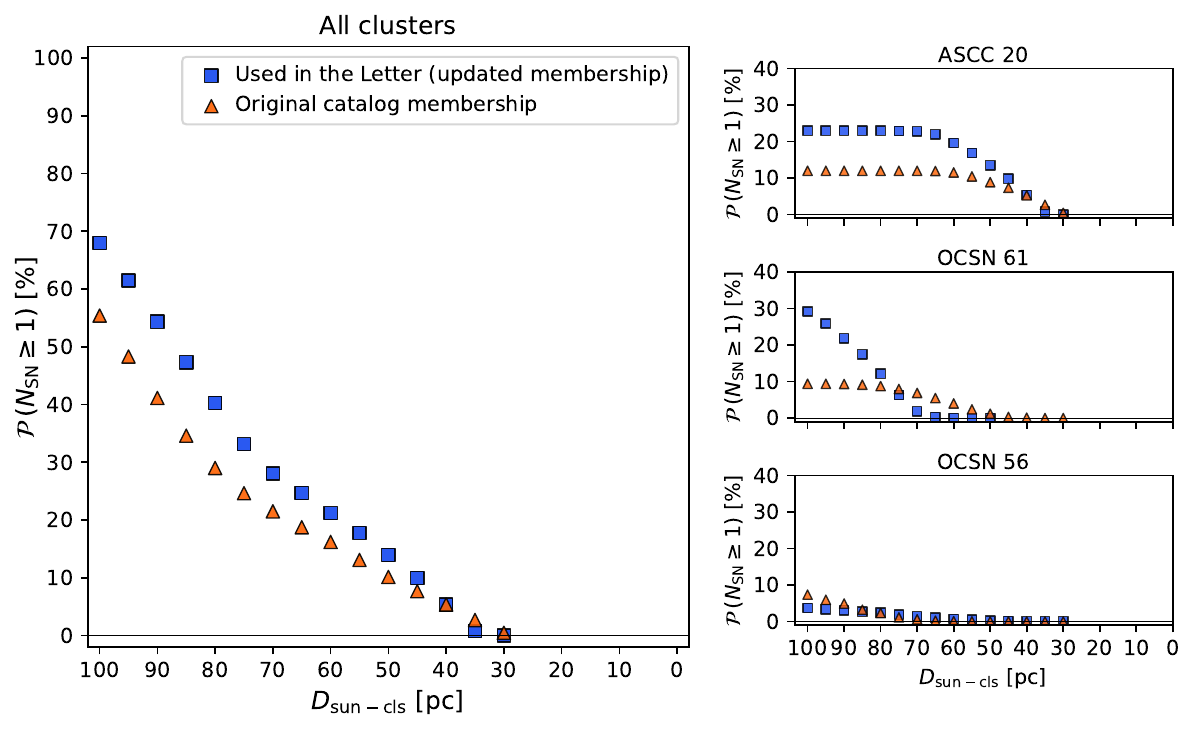}
     \caption{Comparison of the probability of having at least one SN event as a function of distance, between the case where the Orion clusters' updated membership lists are used (main analysis; blue squares) and the case using the original memberships from \citet{Hunt2023} (orange triangles). The left panel shows the total probability from all clusters, while the right panel shows the probabilities for ASCC\,20 (top), OCSN\,61 (middle), and OCSN\,56 (bottom).}
     \label{fig_app:probAtLeast1SN_membershipList}
\end{figure*}

Here, to test the impact of the membership lists, we recomputed the probability of a close SN within the $^{10}$Be anomaly using the original cluster’s memberships as provided by \citet{Hunt2023}. Following the same procedure described in Sect.~\ref{sect:methods}, we determined the cluster ages, applied corrections for mass incompleteness, and estimated both the time and distance to the Solar System of the SNe. The results of this analysis are presented in Fig.~\ref{fig_app:probAtLeast1SN_membershipList} and in Table~\ref{table_app:comparison_OrionMembership}.

We found that:
\begin{itemize}
    \item among the Orion’s clusters, ASCC\,20, OCSN\,56, and OCSN\,61 still represent the main contributors to the SN probability;
    \item for ASCC\,20 and OCSN\,61, the probabilities are lower than those reported in the main analysis due to the smaller number of original stellar members in these clusters. Nevertheless, they remain the main contributors to the total SN probability;
    \item the contribution of ASCC\,20 to the SN probability decreases from 23.0\% to 12.0\% at 100 pc. However, it still dominates at distances below 50 pc, with a probability of 8.9\% (compared to 13.5\% in the main analysis);
    \item the contribution of OCSN\,61 shows the largest decrease, from 29\% to 9.4\% at 100 pc. With the original membership list, however, its contribution begins at a closer distance (about 50 pc instead of 70 pc). This difference arises because the cluster’s velocity and position change slightly when considering the original members rather than the updated ones;
    \item the contribution of OCSN\,56 increases, since its original number of stellar members is greater than in the main analysis. Its probability rises from 3.7\% to 7.3\% at 100 pc;
    \item the total SN probability as estimated from all the clusters decreases from 68.0\% to 55.4\% at 100 pc, and from 13.9\% to 10.1\% at 50 pc. For more details, see Table~\ref{table_app:comparison_OrionMembership}.
\end{itemize}
From this analysis, we conclude that the probability of an SN during the $^{10}$Be anomaly is lower than that found in our main analysis, but still high enough to consider this scenario possible.

\begin{table*}[h!]
\caption{Numerical values of the probability of at least one SN event for the six Orion clusters, comparing the case of updated memberships (as adopted in this paper) with the case where the original memberships are considered.}
\label{table_app:comparison_OrionMembership}    
\centering                        
\resizebox{\textwidth}{!}{%
\begin{tabular}{lccccccccccccccc}
\hline\hline              
Name & 100\,pc & 95\,pc & 90\,pc & 85\,pc & 80\,pc & 75\,pc & 70\,pc & 65\,pc & 60\,pc & 55\,pc & 50\,pc & 45\,pc & 40\,pc & 35\,pc & 30\,pc \\
\hline                      
All clusters  & 68.0 & 61.5 & 54.3 & 47.3 & 40.3 & 33.2 & 28.0 & 24.7 & 21.2 & 17.8 & 13.9 & 10.0 & 5.4 & 0.8 & 0.0 \\
All clusters* & 55.4 & 48.3 & 41.1 & 34.6 & 29.0 & 24.6 & 21.5 & 18.8 & 16.2 & 13.1 & 10.1 & 7.7 & 5.4 & 2.7 & 0.5 \\
\hline 
ASCC\,19  & 0.0 & 0.0 & 0.0 & 0.0 & 0.0 & 0.0 & 0.0 & 0.0 & 0.0 & 0.0 & 0.0 & 0.0 & - & - & - \\
ASCC\,19* & 0.0 & 0.0 & 0.0 & 0.0 & 0.0 & 0.0 & 0.0 & 0.0 & 0.0 & 0.0 & 0.0 & 0.0 & 0.0 & - & - \\
\hline 
ASCC\,20  & 23.0 & 23.0 & 23.0 & 23.0 & 23.0 & 22.9 & 22.8 & 22.0 & 19.6 & 16.9 & 13.5 & 9.8 & 5.3 & 0.8 & 0.0 \\
ASCC\,20* & 12.0 & 12.0 & 12.0 & 12.0 & 12.0 & 12.0 & 12.0 & 11.9 & 11.5 & 10.4 & 8.9 & 7.3 & 5.3 & 2.7 & 0.5 \\
\hline 
OCSN\,56  & 3.7 & 3.4 & 3.1 & 2.7 & 2.4 & 1.8 & 1.4 & 1.0 & 0.7 & 0.5 & 0.2 & 0.1 & 0.1 & 0.0 & 0.0 \\
OCSN\,56* & 7.3 & 5.9 & 4.9 & 3.2 & 2.3 & 1.1 & 0.6 & 0.2 & 0.1 & 0.0 & 0.0 & 0.0 & 0.0 & 0.0 & 0.0 \\
\hline 
OCSN\,61  & 29.2 & 25.9 & 21.8 & 17.5 & 12.2 & 6.3 & 1.9 & 0.3 & 0.0 & 0.0 & 0.0 & - & - & - & - \\
OCSN\,61* & 9.4 & 9.4 & 9.3 & 9.1 & 8.7 & 8.0 & 6.9 & 5.5 & 4.0 & 2.4 & 1.2 & 0.3 & 0.1 & 0.0 & 0.0 \\
\hline 
OCSN\,65  & 0.9 & 0.0 & 0.0 & 0.0 & 0.0 & 0.0 & 0.0 & 0.0 & - & - & - & - & - & - & - \\
OCSN\,65* & 1.6 & 0.8 & 0.7 & 0.4 & 0.1 & 0.1 & 0.0 & 0.0 & - & - & - & - & - & - & - \\
\hline 
Theia\,13  & 0.0 & 0.0 & 0.0 & 0.0 & 0.0 & 0.0 & 0.0 & 0.0 & 0.0 & 0.0 & 0.0 & 0.0 & 0.0 & 0.0 & 0.0 \\
Theia\,13* & 0.0 & 0.0 & 0.0 & 0.0 & 0.0 & 0.0 & 0.0 & 0.0 & 0.0 & 0.0 & 0.0 & 0.0 & 0.0 & 0.0 & 0.0 \\
\hline                                 
\end{tabular}
}
\tablefoot{* indicates cases in which the membership list corresponds to the one provided by the \citet{Hunt2023} catalog.}
\end{table*}

\end{appendix}
\end{document}